\documentclass[aps,twocolumn,reprint,superscriptaddress]{revtex4-1}

\usepackage[T1]{fontenc}
\usepackage[latin9]{luainputenc}
\usepackage{units}
\usepackage{amsmath}
\usepackage{amssymb}
\usepackage{graphicx}
\PassOptionsToPackage{normalem}{ulem}
\usepackage{ulem}
\makeatletter
\numberwithin{equation}{section}
\usepackage{hyperref}
\usepackage{mathtools}
\hypersetup{
  colorlinks   = true, 
  urlcolor     = blue, 
  linkcolor    = blue, 
  citecolor   = blue
}
\bibpunct{[}{]}{,}{n}{}{}
\usepackage{natbib}
\usepackage{url}
\usepackage{tikz}
\usepackage{xifthen}
\usepackage{tikz-cd}
\usetikzlibrary{decorations.markings}
\usepackage{amsmath,amssymb}
\renewcommand*{\theequation}{\arabic{equation}}
\usepackage[caption=false]{subfig}

\begin{document}
\title{ Localization and oscillations of Majorana fermions in a two-dimensional electron gas coupled with $d$-wave superconductors  }

\author{L. Ortiz}
\affiliation{Departamento de F\'{\i}sica Te\'orica I, Universidad Complutense, 28040 Madrid, Spain}
\author{S. Varona}
\affiliation{Departamento de F\'{\i}sica Te\'orica I, Universidad Complutense, 28040 Madrid, Spain}
\author{O. Viyuela}
\affiliation{Department of Physics, Harvard University, Cambridge, MA 02318, USA}
\affiliation{Department of Physics, Massachusetts Institute of Technology, Cambridge, MA 02139, USA}
\author{M.A. Martin-Delgado}
\affiliation{Departamento de F\'{\i}sica Te\'orica I, Universidad Complutense, 28040 Madrid, Spain}

\vspace{-3.5cm}

\begin{abstract}
We study the localization and oscillation properties of the Majorana fermions that arise in a two-dimensional electron gas (2DEG) with spin-orbit coupling (SOC) and a Zeeman field coupled with a $d$-wave superconductor. Despite the angular dependence of the $d$-wave pairing, localization and oscillation properties are found to be similar to the ones seen in conventional $s$-wave superconductors. In addition, we study a microscopic lattice version of the previous system that can be characterized by a topological invariant. We derive its real space representation that involves nearest and next-to-nearest-neighbors pairing. Finally, we show that the emerging chiral Majorana fermions are indeed robust against static disorder. 
This analysis has potential applications to quantum simulations and experiments in high-$T_c$ superconductors.
\end{abstract}


\maketitle

\section{Introduction}\label{sec:introduction}

The idea of a fermionic particle being precisely its own antiparticle has been puzzling  physicists for generations. These exotic particles were hypothesized by  Majorana \cite{Majorana_1st} and have been thoroughly studied in high energy physics as a possible solution to the intriguing nature of neutrinos and dark matter \cite{Wilczek_2009}. However, the detection of Majorana fermions  had remained elusive until they were introduced as quasiparticles in certain condensed matter systems. A series of experiments has claimed the observation of signatures of Majorana states \cite{Das2012,Mourik1003,doi:10.1021/nl303758w,nat_charlie_markus,He294,PhysRevLett.116.257003}.

Roughly, the condensed matter version of Majorana fermions constitutes \textit{half} of a usual fermion, i.e. an ordinary fermion is a superposition of two Majorana modes which can be  separated by arbitrary distance. The resulting state is highly delocalized and robust against local perturbations. Moreover, Majorana states exhibit novel statistics: they are non-abelian anyons. The two latter features draw the attention of  the quantum computation community.  Braiding Majorana fermions provides a method for realizing topological quantum computation \cite{KITAEV20032,1367-2630-12-8-083039,10.1038_nphys1915,RevModPhys.80.1083}. Novel methods combining Majorana physics with topological error correction have also appeared recently \cite{arXiv:1704.01589,arXiv:1708.05012,arXiv:1709.02318,PhysRevLett.97.180501,PhysRevLett.98.160502}.

Correspondingly to these unusual properties, an increasing interest on how to get and manipulate Majorana fermions has grown up. The appearance  of Majorana states was predicted in a system with odd superconducting pairing \cite{1063-7869-44-10S-S29}. Since odd superconducting pairing has not been  found in Nature, the proposal seemed to be unrealistic.  A few years later, a remarkable idea to induce topological superconductivity at the surface of a topological insulator by means of proximity effect made it feasible \cite{PhysRevLett.100.096407}.

Alternatively, approaching a semiconductor nanowire, with spin-orbit coupling and subject to a magnetic field, to the surface of a superconductor induces an odd superconducting pairing among the electrons in the semiconductor \cite{PhysRevLett.105.077001,PhysRevLett.105.177002}. The resulting phase is topological and has Majorana quasiparticles.

Initially, various  experimental setups using $s$-wave superconductors were proposed to host Majorana fermions \cite{PhysRevLett.104.040502, AliceaPhysRevB.81.125318}. Recent experiments confirm the success of the experimental proposal \cite{arXiv:1706.05163, arXiv:1706.04573}. It is also possible to obtain topological superconductivity by depositing magnetic adatoms on top of a conventional $s$-wave superconductor \cite{PhysRevB.88.155420,PhysRevB.89.180505,Nadj-Perge602,PhysRevB.93.094508,PhysRevB.94.125121,Li2016,PhysRevB.94.060505,arXiv:1707.02326}, where signatures of Majorana modes have been seen.

Additionally, high-$T_c$ superconductors were suggested to induce  topological superconductivity \cite{PhysRevB.87.014504,linder_2010,doi:10.1143/JPSJ.81.011013}. The motivation to study these type of superconductors comes in two directions. First, the induced superconducting gap is proportional to the gap in the original superconductor but reduced by a factor, as a consequence of proximity effect. Since the superconducting gap is larger for high-$T_c$ superconductors, the induced gap becomes wider. Second, high-$T_c$ superconductors show anisotropic pairing. Therefore, they induce different pairing depending on the orientation of the  sample. 

 Since high-$T_c$ superconductors  are an instance of a $d$-wave pairing , one may wonder what would happen when a $d$-wave superconductor induces a superconducting gap in a two-dimensional electron gas (2DEG) with spin-orbit coupling and a Zeeman field. A realization of a 2DEG could be a semiconductor even though the aim of this work is to provide a general framework which  can be applied also to other schemes \cite{arXiv:1707.08130} and include quantum simulation in the pathway. 

Considering a host $d$-wave superconductor, we analyze the interesting features of the new Majorana fermions comparing them to the Majorana bound states induced by $s$-wave superconductors. To accomplish this task, we develop a phenomenological model using a $d$-wave superconductor as a parent Hamiltonian to induce superconductivity. As a result, we get an effective pairing which has $f$-wave symmetry $(l=3)$, in contrast to the effective $p$-wave symmetry $(l=1)$ that appears for a parent $s$-wave superconductor.  

$d$-wave superconductors act differently from $s$-wave superconductors in two fundamental ways: 1) $d$-wave pairing shows an angular dependence. As a consequence, $d$-wave superconductors present a richer phenomenology with respect to $s$-wave superconductors. They can induce a $p$-wave paring and also a novel $f$-wave pairing depending on the orientation of the superconductor \cite{PhysRevB.87.014504}. We focus our study on the latter case. Majorana fermions created present clearly defined edge localization, despite the angular dependence of $d$-wave pairing and its expected larger correlation length. We also show that the frecuency of oscillations of Majorana fermions for $d$-wave and $s$-wave superconductors are indeed very similar for a wide range of parameters. 2) Since $d$-wave pairing has nodal lines, where the superconducting gap is zero  \cite{PhysRevLett.105.217001,PhysRevB.93.180501,0953-8984-27-24-243201}, Majorana and nodal states coexist. As a consequence, nodal states appear in the system where the gap in the effective model closes, similarly to what happens for Majorana states.

Furthermore, we study a microscopic lattice Hamiltonian that comprises the previous effective model. We characterize the phase diagram using the parity of the Chern number, which is a well-defined topological invariant even for nodal systems. In addition, we consider the effect of static disorder to prove the robustness of the propagating Majorana modes. 

The paper is organized as follows: In Section \ref{sec:continuous}, a novel superconducting Hamiltonian with effective $f$-wave pairing is derived. We use this simplified Hamiltonian model to calculate analytically the wave function of Majorana fermions in Section \ref{sec:results_1}.  Moreover, we exhaustively study the properties of Majorana fermions arising from this induced $f$-wave pairing.  In Section \ref{sec:lattice}, we study a lattice version of the previous effective model. In particular, we obtain a microscopic model in real space and define the topological invariant for this nodal system. The robustness of Majorana states against disorder is also discussed. Detailed analytic calculations on how to obtain the wave function of the Majorana bound states are explained in Appendix \ref{app:analytical}.

\section{Phenomenological Hamiltonian}\label{sec:continuous}

In this Section, we study the problem of a 2DEG with strong spin-orbit interaction and Zeeman field; as well as an induced $d$-wave superconducting pairing mechanism. There are several physical platforms that can realize this model such as: (i) a planar semiconductor approximated to a high-$T_c$ superconductor \cite{PhysRevLett.104.040502}, (ii) a $d$-wave superconductor with intrinsically strong spin-orbit interaction \cite{PhysRevLett.105.217001}, (iii) cold atoms simulation of $d$-wave superconductors \cite{Zoller}, where the spin-orbit interaction can be also induced by laser \cite{Lukin}.

Any of the above proposals requires three key ingredients: Spin-Orbit Coupling (SOC), an strong Zeeman field and a parent superconductor. The Zeeman field is introduced perpendicular to the semiconductor plane, as shown in Fig. \ref{fig:structure}. In certain parameter regimes, the Hamiltonian presents an effective spin-triplet pairing symmetry with propagating Majorana states at edges, similar to $s$-wave parent superconductors. 
However, we show that the localization, oscillation and stability properties of these new Majorana modes are very similar when the underlying parent symmetry of the superconductor is $d$-wave, despite the angular dependence that the superconducting pairing exhibits. In what follows we present both analytic and numerical results supporting these claims.

\subsection{Derivation of the Hamiltonian}

We begin by considering a 2DEG with SOC \cite{AliceaPhysRevB.81.125318}. Crucially, the SOC breaks the spin degeneracy of the 2DEG bands. Since we will eventually include superconductivity, we already embed the particle-hole structure in the Hamiltonian. To this end, we use the following Nambu spinor basis in momentum space, $
\Psi^{\dagger} (\boldsymbol{k})=\left(\psi^{\dagger}_\uparrow(\boldsymbol{k}),\psi^{\dagger}_\downarrow(\boldsymbol{k}),\psi_\downarrow(-\boldsymbol{k}),-\psi_\uparrow(-\boldsymbol{k})
\right)$, where $\psi (\psi^{\dagger})$ are annihilation (creation) operators satisfying the fermionic anti-commutation relations. The Hamiltonian reads
\begin{align}
\mathcal{H}_{\alpha}=\frac{1}{2}\int\;\text{d}^2\boldsymbol{k}\;\; \Psi^{\dagger}(\boldsymbol{k}) H_{\alpha}(\boldsymbol{k})\Psi(\boldsymbol{k}),
\end{align}
with
\begin{equation}\label{eq:2DEG}
H_{\alpha}(\boldsymbol{k})=\left(\frac{k^{2}}{2m}-\mu\right)\tau_{z}\otimes\mathbb{I}_\sigma + \alpha \;\tau_{z}\otimes\Big(k_{y}\ \sigma_{x}-k_{x}\ \sigma_{y}\Big), 
\end{equation}

\noindent where $k_x (k_y)$ is the crystalline momentum in the $x(y)-$direction, $k^2=k_x^2+k_y^2$, $m$ is the effective mass of the electron in the material, $\mu$ is the chemical potential,  $\alpha$ is the Rashba SOC strength and $\sigma_{i}$ and $\tau_i$ are Pauli matrices acting on spin and particle-hole space respectively. As a result, Hamiltonian $H_{\alpha}(\boldsymbol{k})$ is a $4\times4$ matrix.

Next, we include a Zeeman field perpendicular to the 2DEG plane to open a gap between the spin-up and the spin-down bands:
 \begin{align}\label{eq:h_semi}
&H_{V}=V\;\mathbb{I}_\tau\otimes\sigma_{z}.
\end{align}
The Zeeman field could be generated by a ferromagnetic insulator or by a magnetic field. Since the field was chosen to be perpendicular to the plane, using a magnetic field  would cause orbital effects which are neglected in Eq.\eqref{eq:h_semi}.

The corresponding energy dispersion relations for the particle bands are
 \begin{equation}
  E_{\pm}(\boldsymbol{k})=\frac{k^2}{2m}-\mu\pm \sqrt{V^2+\alpha^2k^2}.
 \end{equation}
where $\pm$ denotes the upper and lower bands respectively. At $\boldsymbol{k}=0$ the separation between the two is $2\left|V\right|$. If $\mu$ is placed inside the gap, $|\mu|<|V|$, only the lower band is occupied. This is necessary to reach the spinless regime.

By placing a superconductor on top of the 2DEG, it is possible to induce superconductivity through proximity effect. Provided we assume spin-singlet pairing, the induced Hamiltonian is given by 

\begin{align}\label{eq:h_pairing}
H_{\Delta}(\boldsymbol{k})=\Delta(\boldsymbol{k})\ \tau_{x}\otimes\mathbb{I}_{\sigma},
\end{align}
where $\Delta(\boldsymbol{k})$ is the induced pairing amplitude, considered real throughout the paper.

Combining all terms from Eqs.~\eqref{eq:2DEG}, \eqref{eq:h_semi} and \eqref{eq:h_pairing}, we find the final Hamiltonian with induced superconducting pairing
\begin{equation}
H=H_\alpha(\boldsymbol{k})+H_V+H_\Delta(\boldsymbol{k}).
\label{Htot}
\end{equation}

In order to further simplify the above Hamiltonian, let us express it in the diagonal basis of $H_\alpha (\boldsymbol{k})+H_V$ with Rashba coupling and Zeeman field only:

 \begin{equation}
H=\left(\begin{array}{cccc}
E_{+} & 0 & \Delta_{+-} & \Delta_{++}\\
0 & E_{-} & \Delta_{--} & \Delta_{+-}\\
\Delta_{+-}^{*} & \Delta_{--}^{*} & -E_{-} & 0\\
\Delta_{++}^{*} & \Delta_{+-}^{*} & 0 & -E_{+}
\end{array}\right),\label{eq: H 2D perpendicular B alicea}
\end{equation}
where
\begin{eqnarray}
\Delta_{--}(\boldsymbol{k})&=&\frac{-\alpha k\Delta(\boldsymbol{k})}{\sqrt{V^{2}+\alpha^{2}k^{2}}}\frac{-ik_{x}+k_{y}}{k},\label{D--}\\
\Delta_{++}(\boldsymbol{k})&=&\frac{-\alpha k\Delta(\boldsymbol{k})}{\sqrt{V^{2}+\alpha^{2}k^{2}}}\frac{ik_{x}+k_{y}}{k},\label{D++}\\
\Delta_{+-}(\boldsymbol{k})&=&\frac{-V\Delta(\boldsymbol{k})}{\sqrt{V^{2}+\alpha^{2}k^{2}}}.\label{D+-}
\end{eqnarray}
Therefore, the originally parent  pairing $\Delta(\boldsymbol{k})$ has generated an effective intraband $\Delta_{++}$, $\Delta_{--}$ and interband $\Delta_{+-}$ pairing.
If the interband coupling $\Delta_{+-}$, which is of the order of $\Delta(\boldsymbol{k})$, is much smaller than the separation between the two particle bands $\approx\left|V\right|$, i.e., $\left|V\right|\gg \Delta(\boldsymbol{k})$,  we can neglect the upper unoccupied band. As a result, we focus on the $2\times2$ effective Hamiltonian given by the lower bands. In this limit, the intraband terms couple electrons with the same spin, i.e. spin-triplet pairing, reaching the spinless regime. This is essential to have Majorana bound states, since creation and annihilation operators for Majorana quasiparticles must be equal to each other in order to fulfill the condition that Majorana fermions are their own antiparticles. 

 For $d$-wave pairing symmetry, the amplitude is given by
\begin{equation}\label{eq:d_pairing}
\Delta(\boldsymbol{k})=\frac{\Delta_{d}}{k^{2}_F}\left(k_{x}^{2}-k_{y}^{2}\right),
\end{equation}
where $k_F$ is the Fermi momentum in the 2DEG.  It is important to highlight that the $d$-wave pairing amplitude, unlike the constant $s$-wave pairing, depends on the azimuth angle, $\theta_{\boldsymbol{k}}$, since $k_x^{2}-k_y^{2}=k^{2} \cos(2\theta_{\boldsymbol{k}})$. The momentum parallel to the interface is conserved and therefore, the induced pairing for $d$-superconductors takes the form given in  Eq. \eqref{eq:d_pairing} \cite{PhysRevB.87.014504,1367-2630-19-4-043026,Zareapour:2012aa,Kashiwaya2000,PhysRevLett.74.3451}. Note that for an $s$-wave parent superconductor where $\Delta(\boldsymbol{k})=\Delta_s$ is constant, the above condition, $\left|V\right|\gg \Delta(\boldsymbol{k})$,  is more restrictive than for a $d$-wave parent superconductor, where $\left|V\right|\gg \Delta(\boldsymbol{k})$ is automatically satisfied for $\boldsymbol{k}\sim0$, unlike the $s$-wave case. We stress that the continuum theory given by Hamiltonian $\eqref{Htot}$ is strictly valid in the vicinity of the $\Gamma$ point $\boldsymbol{k}\approx0$. In this case, the $d$-wave superconductor is itself a
gapless system and supports flat bands at the edge.  The effect of these modes on the 2DEG depends on the particular details of the setup and are considered negligible for the present study. 

\newcommand*\xvar{8}
\newcommand*\yvarSM{2}
\newcommand*\yvarSC{3}
\newcommand*\zvar{5}
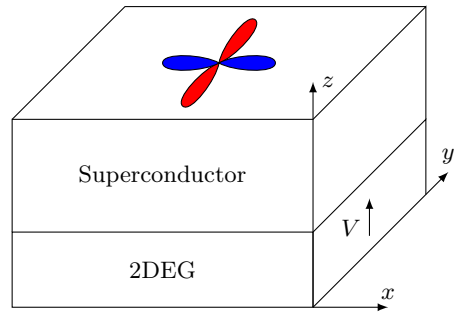
\begin{figure}
\begin{centering}
\begin{tikzpicture}[x=0.5cm,y=0.5cm,z=0.3cm,>=stealth,mylober/.pic = {\begin{scope}[xscale=.5]
         \path (0,0) coordinate(A) (0,3) coordinate(B);
         \draw[postaction={decorate},fill=red] (A)to[out=40,in=0](B) to[out=180,in=140](A);
         \end{scope}},%
         mylobeb/.pic= {\begin{scope}[xscale=.5]
         \path (0,0) coordinate(A) (0,3) coordinate(B);
         \draw[postaction={decorate},fill=blue] (A)to[out=40,in=0](B) to[out=180,in=140](A);
         \end{scope}},
         >=latex,%
         ]
\draw[->] (xyz cs:x=0) -- (xyz cs:x=2) node[above] {$x$};
\draw[->] (xyz cs:y=-0) -- (xyz cs:y=\yvarSM*1.2+\yvarSC*1.2) node[right] {$z$};
\draw[->] (xyz cs:z=0) -- (xyz cs:z=\zvar*1.2) node[above] {$y$};

\draw (xyz cs:x=-\xvar) -- +(0,\yvarSM) --  (xyz cs:y=\yvarSM) --  +(0,-\yvarSM)  -- cycle;
\draw (xyz cs:y=\yvarSM) -- +(xyz cs:z=\zvar) -- (xyz cs:z=\zvar) ;
\draw (xyz cs:y=\yvarSM,z=\zvar) -- +(0,\yvarSC) -- (0,\yvarSC+\yvarSM) -- +(-\xvar,0) --(-\xvar,\yvarSM);
\draw (-\xvar,\yvarSM+\yvarSC)--+(xyz cs:z=\zvar) -- (xyz cs:y=\yvarSM+\yvarSC,z=\zvar) ;

\node[align=center] at (-\xvar/2,\yvarSM/2)  {2DEG};
\node[align=center] at (-\xvar/2,\yvarSM+\yvarSC/2)  {Superconductor};

 \draw pic at (xyz cs:x=-\xvar/2,y=\yvarSM+\yvarSC,z=\zvar/2) [rotate=140,scale=.5 ] {mylober};
 \draw pic at (xyz cs:x=-\xvar/2,y=\yvarSM+\yvarSC,z=\zvar/2) [rotate=320,scale=.5 ] {mylober};
 \draw pic at (xyz cs:x=-\xvar/2,y=\yvarSM+\yvarSC,z=\zvar/2) [rotate=90,scale=.5 ] {mylobeb};
 \draw pic at (xyz cs:x=-\xvar/2,y=\yvarSM+\yvarSC,z=\zvar/2) [rotate=270,scale=.5 ] {mylobeb};

 \draw[->] (xyz cs:y=.2*\yvarSM,z=\zvar/2) -- (xyz cs:y=\yvarSM*.7,z=\zvar/2);
 \node[left] at (xyz cs:y=.5*\yvarSM*0.6,z=\zvar/2) {$V$};

\end{tikzpicture}
\end{centering}
\caption{Orientation of the $d$-wave superconductor with respect to the 2DEG with SOC and Zeeman field.}\label{fig:structure}
\end{figure}

Assuming that the Zeeman field is also much larger than the spin-orbit energy, $\left|V\right|\gg E_{SO}=\frac{1}{2}m\alpha^{2}$, and that we have a parent $d$-wave superconductor, we arrive at the following effective Hamiltonian:
 \begin{align}
H_{\mathrm{eff}}\left(\boldsymbol{k}\right)=\left(\begin{array}{cc}
\frac{k^{2}}{2m}-\mu-|V| &\Delta_f\left(\boldsymbol{k}\right)\\
\Delta^*_f\left(\boldsymbol{k}\right) & -\frac{k^{2}}{2m}+\mu+|V| 
\end{array}\right),\label{eq: Heff approx}
\end{align}
with an induced pairing 
\begin{align}\label{eq:f_pairing}
\Delta_f= \frac{-\alpha\Delta_{d}}{\left|V\right|}\left(-ik_{x}+k_{y}\right)\frac{\left(k_{x}^{2}-k_{y}^{2}\right)}{k^{2}_F},
\end{align}
where $k_F=\sqrt{2m\left(\mu + \left|V\right| \right)}$. If we expand the above equation in polar coordinates to study the orbital symmetries of the pairing, we have: $\Delta_f\sim k^3(e^{3i\theta_k}-e^{-i\theta_k})$. Thus, the resulting pairing has both orbital $p$-wave and $f$-wave symmetries and both form a spin-triplet that allows the existence of Majorana states. From now on, we call the mentioned pairing $f$-wave pairing for simplicity.

We note that the energy gap closes when $\mu+\left|V\right|=0$ signaling a phase transition. For $\mu>-\left|V\right|$ the superconductor is in a topological phase, and in a trivial phase otherwise. Additionally, the energy gap also closes at four nodal points at the Fermi surface when $k_x=\pm k_y=\pm k_F$ \cite{0268-1242-27-12-124003}.
  
It is worth mentioning that the effective $f$-wave pairing in Eq. \eqref{eq:f_pairing} is obtained when the crystallographic orientation of  the $d$-wave parent superconductor with respect to the 2DEG plane is the one shown in Fig. \ref{fig:structure}. Otherwise a different pairing symmetry would be induced \citep{PhysRevB.87.014504}.

\subsection{Majorana wavefunction from $d$-wave superconductors}\label{sec:results_1}

So far we have derived a  effective two-band Hamiltonian, which is simple enough to analytically compute the localization and oscillation properties of the induced Majorana modes. In particular, we would like to study how the wave function of the Majorana fermions is modified due to the inclusion of a parent $d-$wave superconductor, instead of the more commonly studied case with $s$-wave pairing symmetry.

Starting from Hamiltonian in Eq. \eqref{eq: Heff approx}, we assume semi-infinite boundary conditions in the $x$-direction and periodic boundary conditions in the $y$-direction. As mentioned before, we take the pairing amplitude $\Delta(\boldsymbol{k})$ to be real. Majorana zero modes (MZMs) have to fulfill the zero energy condition $H_{\text{eff}}\:\boldsymbol{\psi}=E\boldsymbol{\psi}=0$ at $k_y=0$, giving rise to the following system of differential equations
\begin{equation}
\begin{cases}
\frac{1}{2m}\partial_{x}^{2}\psi_{1}+\frac{\alpha\Delta_{0}}{\left|V\right|}\partial_{x}^{3}\psi_{2}+\left(\mu+\left|V\right|\right)\psi_{1} & =0\\
\frac{1}{2m}\partial_{x}^{2}\psi_{2}+\frac{\alpha\Delta_{0}}{\left|V\right|}\partial_{x}^{3}\psi_{1}+\left(\mu+\left|V\right|\right)\psi_{2} & =0
\end{cases}.
\label{eq: Majorana H_eff system of diff equations}
\end{equation}

The equation above can be easily decoupled using particle-hole symmetry, since $\psi_1$ and $\psi_2$ are related by $\psi_1=-\psi_2$. We are left with a single independent linear differential equation where the third order term comes from the $f$-wave pairing in Eq. \eqref{eq:f_pairing}. Subsequently, a third degree characteristic polynomial is solved to find the solutions. Since we consider a semi-infinite system and we are looking for localized states at the edge, we enforce the boundary conditions: $\psi(x=0)=0$ and $\psi(x=\infty)=0$. The only possible solution for Eq.~\eqref{eq: Majorana H_eff system of diff equations} with these constraints casts the form
\begin{equation}\label{eq:majoranas}
\psi_1\left(x\right)=Ne^{-ux}\sin vx, 
\end{equation}
where $u$ and $v$ are, respectively, the real and imaginary parts of one of the roots of the characteristic polynomial.  Since $u$ and $v$ are the solutions for a $d$-wave parent Hamiltonian, henceforth we will add the subindex $u_d$ and $v_d$ to denote these solutions. For an $s$-wave parent Hamiltonian, the solution for MZMs also takes the form in Eq. \eqref{eq:majoranas}. In this case, the solutions are called $u_s$ and $v_s$. We study first the properties of Majorana wave function coming from a $d$-wave superconductor and we compare the new results with the MZMs induced by an $s$-wave superconductor.

The decay of the MZMs into the bulk is given by $u_d$, and the amplitude of the oscillation of the wave function by $v_d$. The third order differential equation for $\psi_1$ leads to a third order algebraic equation for $u_d$. Making use of the relations between the coefficients of a third order polynomial and its roots, it is possible to find an explicit expression for $u_d$: 
\begin{alignat}{2}
&u_d=	&&-\frac{\left|q\right|}{q}\sqrt{\frac{p}{3}}\cosh\left(\frac{1}{3}\mathrm{arccosh}\left(\frac{3\left|q\right|}{2p}\sqrt{\frac{3}{p}}\right)\right) \nonumber\\
	&\ &&-\frac{1}{3}\left|\frac{\left|V\right|\left(\mu+\left|V \right|\right)}{\alpha\Delta_{d}}\right|, \label{u}
\end{alignat}
where $p$ and $q$ are defined as
\begin{equation}\label{eq:p}
p\coloneqq\frac{1}{3}\left|\frac{V}{2m\alpha\Delta_{d}}\right|^{2},
\end{equation}
\begin{equation}\label{eq:q}
q\coloneqq-\frac{2}{27}\left|\frac{V}{2m\alpha\Delta_{d}}\right|^{3}-\left(\mu+\left|V\right|\right)\left|\frac{V}{\alpha\Delta_{d}}\right|.
\end{equation}
The detailed calculation of these expressions is specified in Appendix \ref{app:analytical}.
\begin{figure}
\begin{center}
\includegraphics[width=0.5\textwidth]{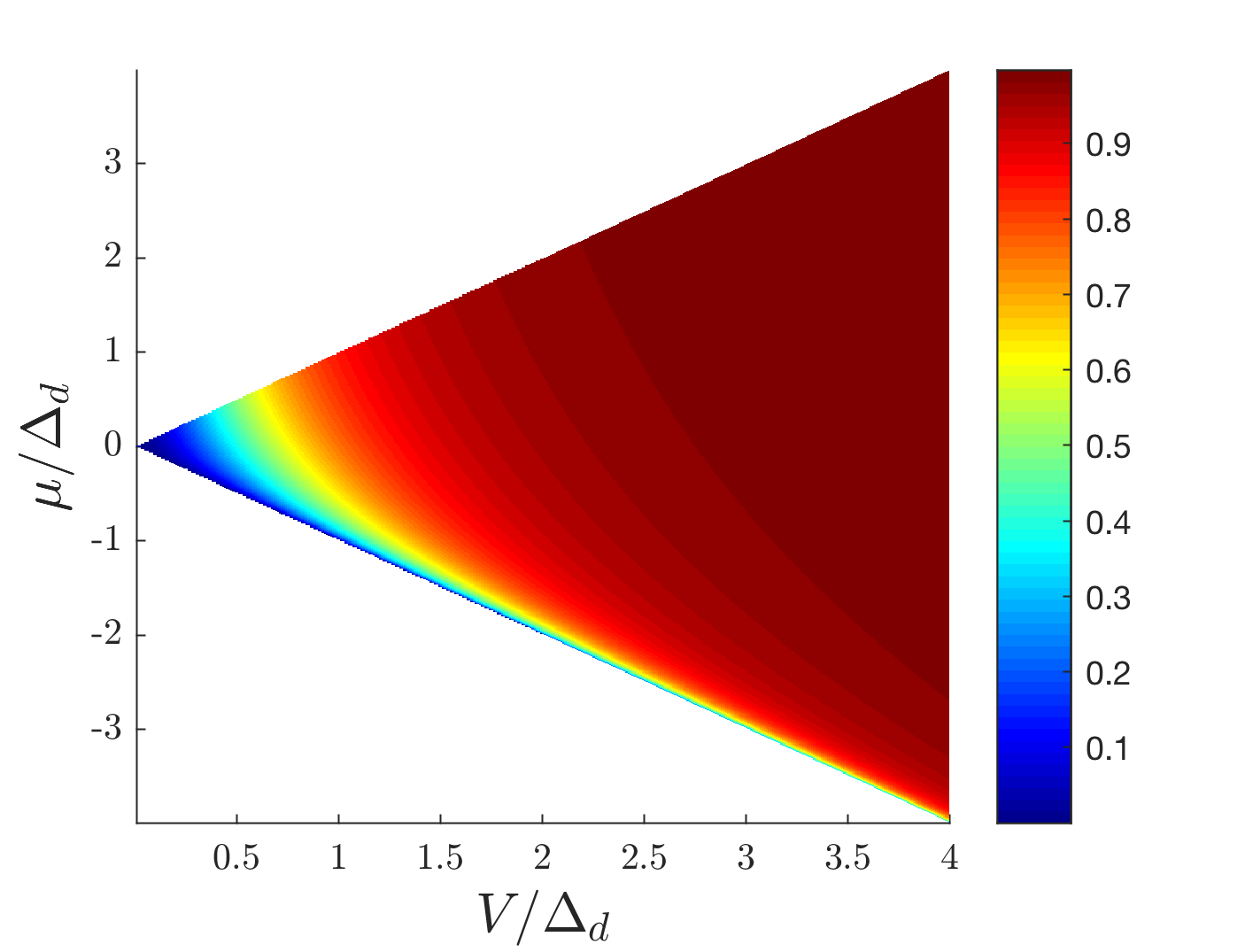}
\caption{Ratio between the exponential decays, $u_d/u_s$, of Majoranas coming from $d$- and $s$-wave  superconductors, respectively, as a function of $V$ and $\mu$.  Only the area with $\left|\mu\right|<\left|V\right|$ is shown. Parameters: $\Delta_s=\Delta_d$ and $E_{SO}=0.05\Delta_d$.}\label{fig: exponential decay ratio mu B}
\end{center}
\end{figure}

Having the analytic expression of the decay of the MZMs, one may wonder why it is important that Majorana fermions remain localized. This is crucial for instance from the point of view of quantum computation, since one reason for the protection of the MZMs is due to their non-local character that result into protection against local perturbations.

The coherence length of the superconductor $\xi$ is inversely proportional to the superconducting gap. Consequently, the larger the gap, the more localized we may expect the MZMs to be. This is true regardless of whether the underlying pairing symmetry is $s$-wave or $d$-wave. The angular dependence for $d$-wave superconductors, Eq.~\eqref{eq:d_pairing}, leads to an effective reduction of the superconducting gap, which implies a larger coherence length $\xi$ on average in the superconductor. As a result, we may intuitively expect a stronger interaction between Majorana fermions at the edges due to this larger coherence length.
An exponentially small gap in the length of the sample opens because of the interaction of the two edge Majoranas. Therefore, edge localization constitutes a figure of merit for the usefulness of MZMs.

Remarkably, we show that for a wide range of values in the system parameters, the effect of the angular dependence of the $d$-wave pairing symmetry is irrelevant and MZMs are as isolated as for an underlying $s$-wave pairing. Furthermore, since experiments show a larger pairing gap for $d$-wave superconductors  $\Delta_d\gg\Delta_s$ \cite{Hashimoto:2014aa,PhysRevB.84.144522,arxiv_1703.03699,2017arXiv170505049L,PhysRevB.93.155402}, the localization of MZMs should be even more pronounced in that case.

\begin{figure}
\includegraphics[width=0.5\textwidth]{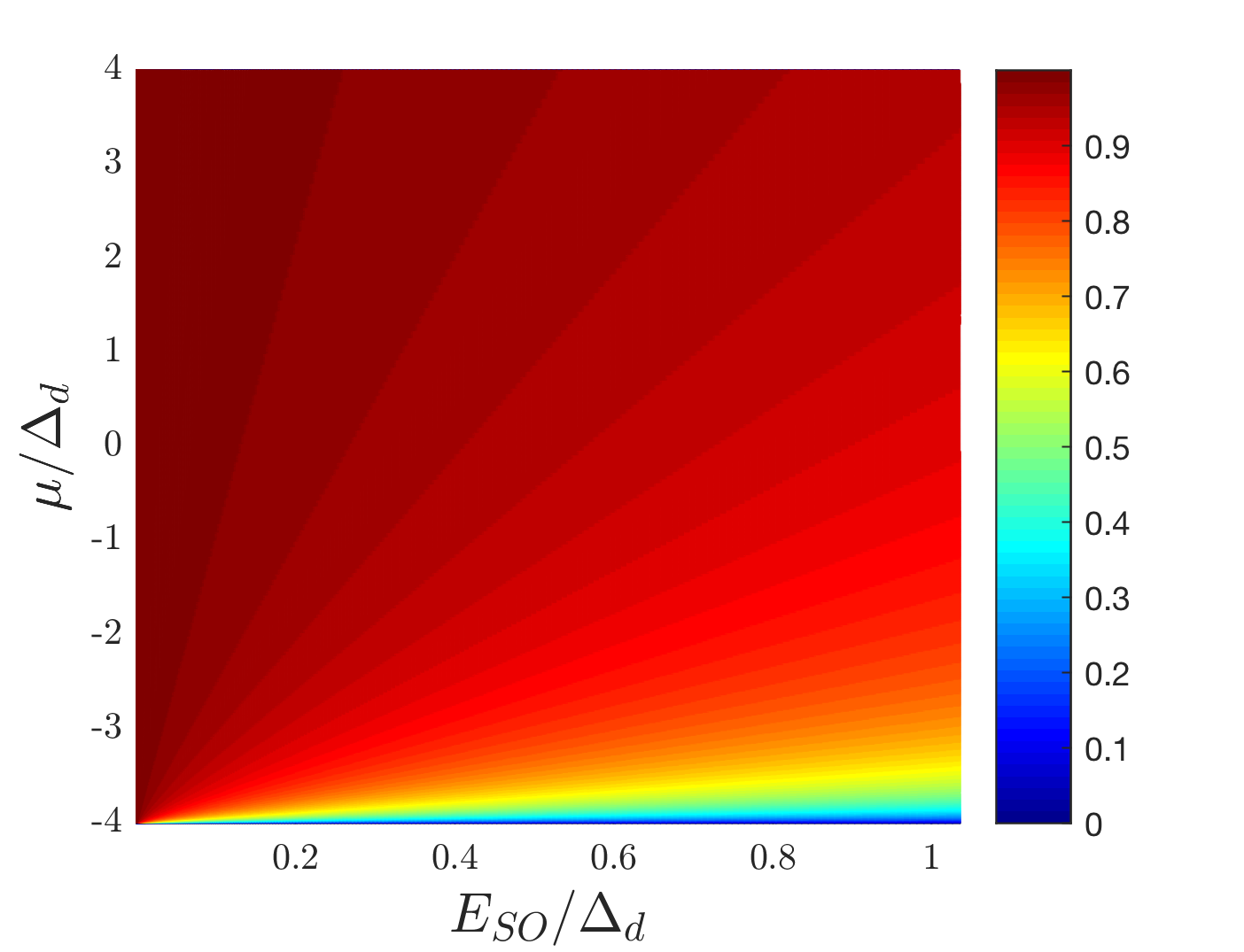}
\caption{Ratio between the localization of MZMs induced by  $d$ and $s$-wave superconductivity, $u_d/u_s$, as a function of $E_{SO}$ and $\mu$. Parameters: $\Delta_s=\Delta_d$ and  $V=4\Delta_d$.}
\label{fig: exponential decay ratio E mu}
\end{figure}

Larger values of $u_d$ mean that MZMs are more localized and decay faster into the bulk. We compare the decay $u_d$ with $u_s=\left|\frac{\alpha\Delta_{s}}{V}\right|m$, where $\left|V\right|\gg\Delta_s,E_{SO}$, in Figs. \ref{fig: exponential decay ratio mu B} and \ref{fig: exponential decay ratio E mu}. Fig. \ref{fig: exponential decay ratio mu B}  shows the ratio $u_d/u_{s}$ as function of $V$ and $\mu$ considering the underlying $s$-wave superconducting gap equal to the $d$-wave one, $\Delta_s=\Delta_d$ in order to isolate and study the influence of the angular dependence solely. Fig. \ref{fig: exponential decay ratio E mu} shows this same ratio as a function of $E_{SO}$ and $\mu$. The ratios between the Hamiltonian parameters are taken to resemble experimental values in semiconductors such as InAs or InSb \cite{PhysRevB.84.144522,2017arXiv170505049L,PhysRevB.93.155402}.

In Fig. \ref{fig: exponential decay ratio mu B} it can be seen that provided the Zeeman energy $V$ is large with respect to $\Delta_d$, the ratio $u_d/u_{s}$ is nearly equal to 1. This implies that the larger $V$ the more similar the $s$-wave and the $d$-wave case become, regardless of the angular dependence of the $d$-wave pairing. 

Nevertheless, there are certain areas where the ratio $u_d/u_s$ decreases, where the localization of MZMs coming from $s$-wave superconductor is much greater than the localization of Majoranas induced by a $d$-wave superconductor. This can be seen in a small band in the lower part of Fig. \ref{fig: exponential decay ratio E mu} where $\mu\sim-V$. Also, in Fig. \ref{fig: exponential decay ratio mu B} the ratio decreases for low values of $V$. However, this latter area is outside the valid regime of our effective model. In Fig. \ref{fig: exponential decay ratio mu B}, we need to have $\left|V\right|\gg \Delta_s$ for $u_s$ (and $\left|V\right|\gg \Delta(\boldsymbol{k})$ for $u_d$) and $|V|\gg E_{SO}$ in Fig. \ref{fig: exponential decay ratio E mu}.

A possible instance of  $d$-wave superconductors corresponds to high-$T_c$ superconductors, where the superconducting gap is one or two orders of magnitude greater than a conventional $s$-wave superconductor \cite{Hashimoto:2014aa,Zareapour:2012aa}. This means that $\Delta_s$ would be much smaller than $\Delta_d$. Considering realistic $\Delta_s$ values, the ratios in Figs. \ref{fig: exponential decay ratio mu B} and \ref{fig: exponential decay ratio E mu} are multiplied by the relation $\Delta_d/\Delta_s$. Thus, MZMs arising from $d$-wave superconductors should be much more localized for high $V$ than their counterparts, induced by $s$-wave superconductors. 

Additionally, the wave function of MZMs, see Eq. \eqref{eq:majoranas}, oscillates at a frequency $v_d$,
\begin{align}
v_d^{2}=\frac{2 (s_1+2u)}{u},
\label{v}
\end{align}
where $s_1=\frac{\left(\mu+\left|V\right|\right)\left|V\right|}{\alpha\Delta_{d}}$. A detailed derivation of Eq. \eqref{v} can be found in Appendix \ref{app:analytical}. The equivalent expression for Majoranas coming from an $s$-wave superconductor is $v_s=\sqrt{2m\left(\mu+\left|V\right|\right)-m^{2}\left|\frac{\alpha\Delta_s}{V}\right|^{2}}$. It is important to remark that $v_s$ can take imaginary values. For $2m\left(\mu+\left|B\right|\right)<m^{2}\left|\frac{\alpha\Delta}{B}\right|^{2}$ the square root is imaginary and the sine of $v_s$ turns into a hyperbolic sine. Therefore, in order to compare $v_s$ and $v_d$, we should consider $v_s=0$ for the range mentioned before. Nevertheless, this occurs when $V/\Delta_s$ is small. Therefore this region with $v_s=0$ is outside the scope of our calculations, since we require $|V|\gg \Delta_s$. Outside these regions, when $\left|V\right|\gg\Delta_s,E_{SO}$ we have $v_s\simeq\sqrt{2m\left(\mu+\left|V\right|\right)}=k_F$. The resulting ratios $v_d/v_s$ have very similar behavior to $u_d/u_s$, taking values close to 1 but always smaller. The larger the Zeeman energy, $V$, the closer this ratio gets to 1. It was mentioned previously that the ratio $u_d/u_s$ is multiplied by a factor $\Delta_d/\Delta_s$ when $\Delta_d\neq\Delta_s$. Contrary to what happens with the decay ratio, $u_d/u_s$, the oscillation ratio does not appreciably change with $\Delta_d/\Delta_s$ and keeps always values close to 1.

In sum, we have proven that the angular dependence of an underlying $d$-wave superconductor has little effect on the localization and oscillation properties of MZMs induced on a 2DEG. Moreover, since $d$-wave superconductors have larger superconducting pairing amplitudes, we may expect MZMs to be more localized than in the $s$-wave parent superconductor case. This fact could have positive implications in current proposals for topological quantum computation using MZMs, since the robustness of the Majorana quasiparticles partly relies on their non-local and edge-localized character.

\section{Microscopic Hamiltonian}\label{sec:lattice}

In Sec. \ref{sec:continuous} we have derived an effective model to study the low energy physics around $\boldsymbol{k}\sim0$. In this section, we define a microscopic lattice model that comprises the effective Hamiltonian previously described in Eq.~\eqref{eq: Heff approx}. Moreover, we calculate a topological invariant which distinguishes between topological and trivial phases, i.e. whether Majorana states exist or not. his microscopic model corresponds to the exotic pairing phenomenologically derived in Sec. \ref{sec:continuous}.

\subsection{Lattice Hamiltonian}

\setcounter{equation}{19}
\renewcommand*{\theequation}{\arabic{equation}}

Assuming that $k$ and $k^2$ terms in Eq.~\eqref{eq: Heff approx} correspond to the lowest order expansion of the trigonometric functions $\sin{k}$ and $\cos{k}$, we can write a lattice Hamiltonian in momentum space that casts the form
\begin{align}
\mathcal{H}_{\mathrm{micro}}=\frac{1}{2}\sum_{\boldsymbol{k}}\left(c_{\boldsymbol{k}}^{\dagger},\ c_{-\boldsymbol{k}}\right)H_{\mathrm{m}}\left(\boldsymbol{k}\right)\left(\begin{array}{c}
c_{\boldsymbol{k}}\\
c_{-\boldsymbol{k}}^{\dagger}
\end{array}\right),
\end{align}
where
\begin{align}
H_{\mathrm{m}}\left(\boldsymbol{k}\right)=\left(\begin{array}{cc}
\epsilon(\boldsymbol{k}) & d(\boldsymbol{k})\\
d^*(\boldsymbol{k}) & -\epsilon(\boldsymbol{k})
\end{array}\right),\label{eq: lattice Hamiltonian}
\end{align}
with $\epsilon\left(\boldsymbol{k}\right)=-2t\left(\cos k_{x}\right.\allowbreak\left.+\cos k_{y}\right)-\tilde{\mu}+4t$, $d\left(\boldsymbol{k}\right)=4i\tilde{\Delta}\left(\sin k_{x}+i\sin k_{y}\right)\left(\cos k_{y}-\cos k_{x}\right)$, $t=\nicefrac{1}{2m}$, $\tilde{\mu}=\mu+\left|V\right|$ and $\tilde{\Delta}=\frac{\alpha\Delta_{d}}{2\left|V\right|k^2_F}$. We recover Eq. \eqref{eq: Heff approx} in the $\boldsymbol{k}\rightarrow0$ limit. The energy bands for the lattice model are given by $E\left(\boldsymbol{k}\right)=\pm\sqrt{\epsilon^{2}\left(\boldsymbol{k}\right)+\left|d\left(\boldsymbol{k}\right)\right|^{2}}$.
The gap vanishes at the points $\left(k_{x},k_{y},\tilde{\mu}\right)=\left(0,0,0\right),\ \left(0,\pi,4t\right),\ \left(\pi,0,4t\right)$ and $\left(\pi,\pi,8t\right)$, suggesting phase transitions. Additionally, there are nodal lines, placed at $k_{x}=\pm k_{y}=\pm k_F$.

\begin{figure}
  \includegraphics[width=\linewidth]{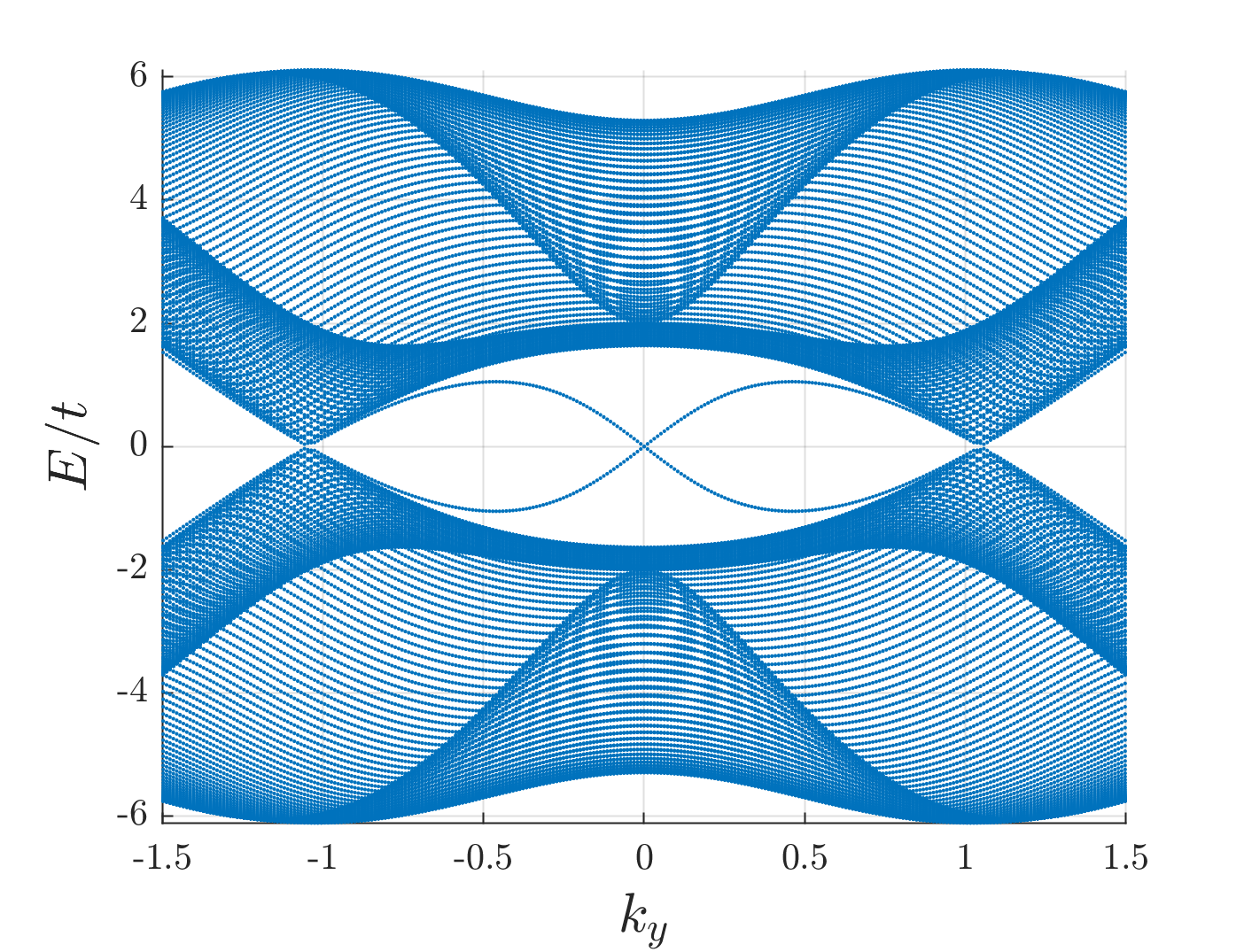}%
\caption{Energy spectra for the $f$-wave lattice model \eqref{eq: lattice Hamiltonian} on a cylindrical geometry. Parameters: $\tilde{\mu}=2t$, $\tilde{\Delta}=t$. Lattice sites in $x$-direction $N$=100. Chiral gapless edge modes can be seen, since the system is topological for these parameters. }
\label{fig: Energy spectrum}
\end{figure}

 Despite the existence of nodal lines that render the system gapless, it is possible to define a topological invariant that distinguishes between non-trivial and trivial phases. The Chern number \cite{kohmoto} calculated for nodeless superconductors is no longer well-defined \cite{PhysRevB.81.220504,PhysRevLett.105.217001}. To define the Chern number in a system with gapless lines like ours, it is necessary to remove the nodal states by adding a small perturbation. Nevertheless, the value of the Chern number is not independent of the perturbation introduced, and only the parity of the Chern number is uniquely defined by this procedure. Thus, this is a well-defined topological invariant even in the presence of bulk gapless excitations. The parity of the Chern number can be computed as
\begin{equation}
\left(-1\right)^{\nu_{\mathrm{Ch}}}=\prod_{\alpha,i=1,2,3,4}\mathrm{sgn}\ E_{\alpha}\left(\Gamma_{i}\right),
\end{equation}
where $E_{\alpha}\left(\boldsymbol{k}\right)$ is the
eigenvalue of Hamiltonian (\ref{eq: lattice Hamiltonian}) for each band $\alpha$. In our particular case, $\alpha$ only takes one single value because the model only has one independent band, due to particle-hole symmetry. $\Gamma_{i}$ are the time-reversal-invariant momenta $\left(0,0\right)$, $\left(0,\pi\right)$, $\left(\pi,0\right)$ and $\left(\pi,\pi\right)$.  Since $d\left(\boldsymbol{k}\right)$ vanishes at time-reversal-invariant momenta we have
\begin{equation}
E_{1}\left(\Gamma_{i}\right)=\epsilon\left(\Gamma_{i}\right)=-2t\left(\cos\Gamma_{i,x}+\cos\Gamma_{i,y}\right)-\tilde{\mu}+4t.
\end{equation}
Applying the definition, the following expression is obtained: 
\begin{align}
\left(-1\right)^{\nu_{\mathrm{Ch}}}=\mathrm{sgn}\left[\left(-\tilde{\mu}\right)\left(-\tilde{\mu}+4t\right)^{2}\left(-\tilde{\mu}+8t\right)\right],
\end{align}
and the parity of the Chern number is $-1$ in the interval $0<\tilde{\mu}<8t$, where the system is in a topological phase. The lower phase boundary $\tilde{\mu}=0$ is in agreement with the results shown in Section \ref{sec:results_1}, where there was a topological phase transition at $\tilde{\mu}=0$. Additionally, the lattice model presents an upper bound for the topological phase at $\tilde{\mu}=8t$ arising from the gap closing at the $M$ point $(k_x,k_y)=(\pi,\pi)$. This feature was not captured in the phenomenological analysis around the $\Gamma$ point $\boldsymbol{k}=0$.

In order to obtain a microscopic model in real space, we employ the inverse Fourier transform:

\begin{align}\label{eq:fourier_transform}
 c_{k_{x},k_{y}}=\frac{1}{L}\sum_{n,m}e^{ink_{x}}e^{imk_{y}}c_{n,m},
 \end{align}
  where $n$($m$) runs over all sites in $x$($y$)-direction and the result of this calculation is:

 \begin{alignat}{1}\label{eq:microscopic_model}
\mathcal{H}_{\mathrm{micro}}= & \sum\nolimits_{m,n}\left\{ -(\tilde{\mu}-4t)c_{m,n}^{\dagger}c_{m,n}-t\left(c_{m+1,n}^{\dagger}c_{m,n}\right.\right.\nonumber \\
 & \left.+c_{m,n}^{\dagger}c_{m+1,n}+c_{m,n+1}^{\dagger}c_{m,n}+c_{m,n}^{\dagger}c_{m,n+1}\right)\nonumber \\
 & +\tilde{\Delta}(c_{m+1,n+1}^{\dagger}c_{m,n}^{\dagger}+c_{m+1,n}^{\dagger}c_{m,n+1}^{\dagger})+\mathrm{H.c.}\nonumber \\
 & -i\tilde{\Delta}(c_{m+1,n+1}^{\dagger}c_{m,n}^{\dagger}+c_{m,n+1}^{\dagger}c_{m+1,n}^{\dagger})+\mathrm{H.c.}\nonumber \\
 & \left.-\tilde{\Delta}(c_{m+2,n}^{\dagger}c_{m,n}^{\dagger})+i\tilde{\Delta}(c_{m,n+2}^{\dagger}c_{m,n}^{\dagger})+\mathrm{H.c.}\right\} .
\end{alignat}
Notably, the pairing in Eq.~\eqref{eq: lattice Hamiltonian}, when transformed from momentum to real space as it is done in Eq.~\eqref{eq:microscopic_model}, involves nearest and next-to-nearest-neighbors interactions. This is in marked contrast to the microscopic model coming from  a host $s$-wave superconductor, since the latter involves only nearest-neighbors interactions. 

We want to study the properties of propagating Majorana states hosted by \eqref{eq:microscopic_model}. Thus, we consider a cylindrical geometry with periodic boundary conditions in the $y$-direction and open boundary conditions in the $x$-direction. In Fig. \ref{fig: Energy spectrum} we depict the energy spectrum for this particular geometry. The propagating Majorana states cross linearly at $k_y=0$ and are separated by a gap from the bulk states.  At $k_y=\pm k_F$ the gap closes again at the Fermi momentum $k_F$, due to the nodal character of the superconducting pairing. 

\subsection{Disorder analysis}\label{sec:results}

We observe the stability of Majorana fermions under static disorder in our lattice model. A random perturbation which depends on the site position modifies slightly the chemical potential. In order to introduce this perturbation, we add a new term to Hamiltonian in \eqref{eq:microscopic_model}, namely:
\begin{align}
\mathcal{H}_{\delta\tilde{\mu}}=\sum_{m,n}\delta\tilde{\mu}_{m,n}c^{\dagger}_{m,n}c_{m,n}.
\end{align}
The coefficients $\delta\tilde{\mu}_{m,n} \in [-\sigma_\mu,\sigma_\mu]$ are picked from a random uniform distribution with zero mean value and width $2\sigma_\mu$. We seek to probe the edge localization of the zero modes. Results show that even in the presence of a random potential, the propagating Majorana modes are robust. Similar studies in odd-frequency $s$-wave pairing show that  Majorana fermions are also robust against disorder \cite{PhysRevB.87.104513, PhysRevLett.98.037003}.

\begin{figure}
\subfloat[\label{fig:f_wave_spectra}]{%
  \includegraphics[width=0.7\linewidth]{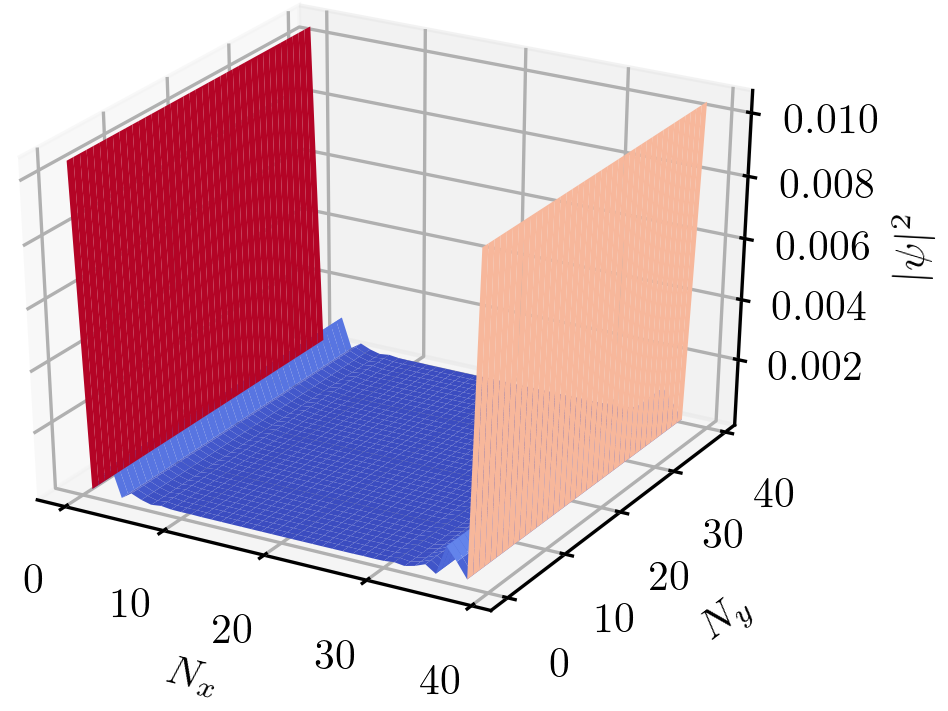}%
}\hfill
\subfloat[\label{fig:f_wave_disorder}]{%
 \includegraphics[width=0.7\linewidth]{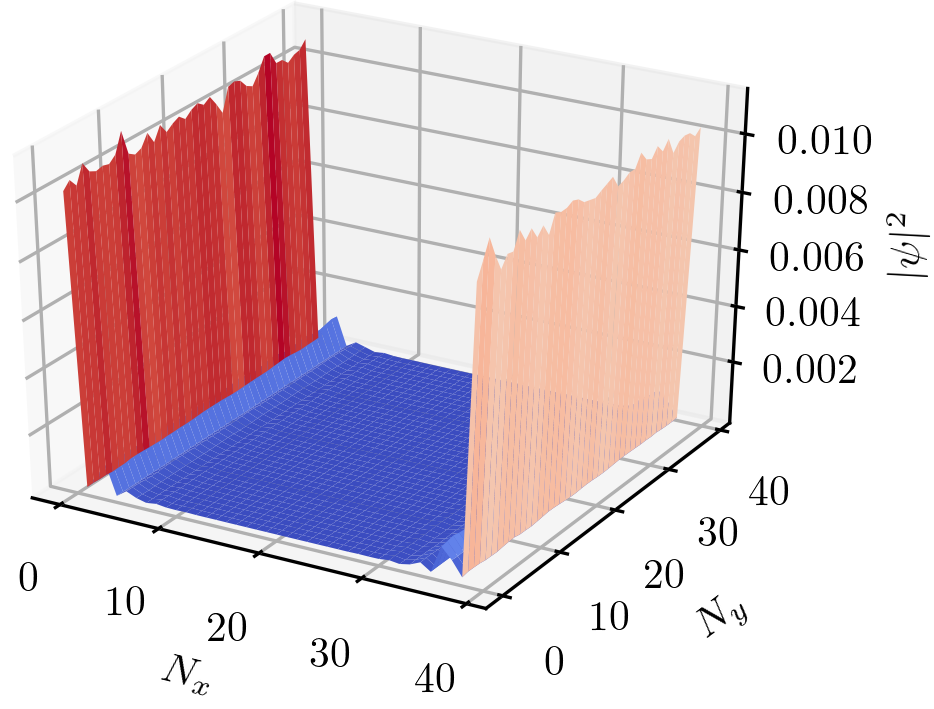}%
}
\caption{Wave functions of the zero energy edge modes on a cylindrical geometry. (a) Shows the wave function of the zero-energy state with no disorder in the system. (b) Depicts the same state with $\sigma_\mu=0.1$.  Parameters: $\tilde{\mu}=2t$, $\tilde{\Delta}=t=1$. Lattice size is $N_x\times N_y=40\times40$.}
\label{fig:Disorder}
\end{figure}

 We introduce static disorder in both $x$- and $y$-direction, considering a cylindrical geometry for our system. Neither $k_x$ nor $k_y$ are good quantum numbers now, since we are breaking translational symmetry. Thus, we calculate the spectrum of the perturbed Hamiltonian and focus on the low energy states. We plot the wave function of zero energy modes to check the localization of MZMs in Fig. \ref{fig:Disorder}.  The results obtained show that even in the presence of  weak static disorder, the edge states remain localized. Moreover, the exponential decay that characterizes MZMs is preserved up to a scale of energies where the static disorder could be treated as a perturbation with respect to the other energies in the system (see Fig. \ref{fig:Disorder_edge} in Appendix  \ref{app:decay}). Majorana fermions may interact with nodal states under certain conditions making the Majoranas less robust \cite{PhysRevLett.105.217001}.

On the other hand, nodal states exposed to static disorder may change their position in momentum space, but cannot be removed. These states appear when the gap closes, $E=\sqrt{\epsilon^{2}\left(\boldsymbol{k}\right)+\left|d\left(\boldsymbol{k}\right)\right|^{2}}=0$, which can only happen if $\epsilon\left(\boldsymbol{k}\right)=d\left(\boldsymbol{k}\right)=0$. $\epsilon\left(\boldsymbol{k}\right)=0$ is the Fermi surface, while $d\left(\boldsymbol{k}\right)=0$ yields the nodal lines $k_x=\pm k_y$. The intersection of the nodal lines and the Fermi surface results in the nodal states. Static disorder introduces a perturbation, $\delta \tilde{\mu}$, which consequently alters the Fermi surface, $\epsilon + \delta \epsilon$, changing the point at which nodal lines cross the surface \cite{PhysRevB.73.214502}.

\section{Conclusions and Outlook }\label{sec:conclusions}
The purpose of this work is to study the properties of emerging Majorana modes in a 2DEG with strong spin-orbit coupling, a Zeeman field and proximity induced $d$-wave superconductivity. Although the angular dependence of $d$-wave superconducting pairing that would intuitively increase the superconducting coherence length, we have remarkably shown that Majorana modes are almost as localized as the ones obtained with a constant $s$-wave pairing amplitude. Moreover, since realistic values of the $d$-wave gap are much greater than the $s$-wave superconducting gap, sharper localization profile is expected for Majorana states induced by the former.

We have also studied a microscopic lattice version of the previous model with an effective $f$-wave pairing. In real space this model involves nearest and next-to-nearest-neighbors interactions. We have computed the phase diagram of this model by means of the parity of the Chern number, a topological invariant that is well-defined even for nodal systems. In addition, we have proven the stability of the propagating Majorana modes against static disorder.

This analysis has direct implications in experiments and proposals with high-$T_c$ superconductors \cite{KITAEV20032}, as well as in cold atoms simulation of $d$-wave superconductors \cite{Zoller}, with Raman-induced spin-orbit coupling \cite{Nadj-Perge602}.

\section*{Acknowledgements}
We thank Debanjan Chowdhury and Liang Fu for helpful discussions. 
We acknowledge financial support from the Spanish MINECO grants FIS2012-33152, FIS2015-67411, and the CAM research consortium QUITEMAD+, Grant No. S2013/ICE-2801. The research of M.A.M.-D. has been supported in part by the U.S. Army Research Office through Grant No. W911N F-14-1-0103.
O.V. thanks Fundaci\'on Rafael del Pino, Fundaci\'on Ram\'on Areces and RCC Harvard. S.V. thanks FPU-MECD Grant. 

\appendix
\setcounter{equation}{0}
\renewcommand*{\theequation}{A\arabic{equation}}
\section{Derivation of  Majorana states from an effective $f$-wave pairing}\label{app:analytical}

Given a cylindrical geometry for Hamiltonian \eqref{eq: Heff approx}, we look for MZMs solutions, that satisfy the equation $H_{\text{eff}}\:\boldsymbol{\psi}=0$ at $k_{y}=0$. This yields the system of differential equations in Eq. \eqref{eq: Majorana H_eff system of diff equations}. Using particle-hole symmetry we can decouple these equations and obtain a single differential equation:
\begin{equation}
\frac{\alpha\Delta_{d}}{\left|V\right|k_{F}^{2}}\partial_{x}^{3}\psi_{1}-\frac{\partial_{x}^{2}}{2m}\psi_{1}-\left(\mu+\left|V\right|\right)\psi_{1}=0,\label{eq: qubic Heff diff eq 1}
\end{equation}
which has a third derivative instead of the first derivative we would find
in the $p$-wave case. Using the ansatz $e^{kx}$ we
obtain the associated
characteristic polynomial of \eqref{eq: qubic Heff diff eq 1}:
\begin{equation}
k^{3}-\frac{\left(\mu+\left|V\right|\right)\left|V\right|}{\alpha\Delta_{d}}k^{2}-\frac{2m\left(\mu+\left|V\right|\right)^{2}\left|V\right|}{\alpha\Delta_{d}}=0.\label{eq:cubic polynomial}
\end{equation}
We will now discuss when does
equation (\ref{eq: qubic Heff diff eq 1}) have Majorana solutions
with boundary conditions $\psi\left(0\right)=\psi\left(\infty\right)=0$.
Polynomial (\ref{eq:cubic polynomial}) may be rewritten as $\left(k-k_{1}\right)\left(k-k_{2}\right)\left(k-k_{3}\right)=k^{3}-s_{1}k^{2}+s_{2}k-s_{3}$,
where $k_{1}$, $k_{2}$ and $k_{3}$ are the roots of the cubic polynomial
and $s_{1}=k_{1}+k_{2}+k_{3}=\frac{\left(\mu+\left|V\right|\right)\left|V\right|}{\alpha\Delta_{0}}$,
$s_{2}=k_{1}k_{2}+k_{2}k_{3}+k_{1}k_{3}=0$ and $s_{3}=k_{1}k_{2}k_{3}=\frac{2m\left(\mu+\left|V\right|\right)^{2}\left|V\right|}{\alpha\Delta_{0}}$.
The discriminant of the cubic equation reads
\begin{equation}
D=-4s_{1}^{3}s_{3}-27s_{3}^{2}.
\end{equation}
$D$ vanishes for $\left(\mu+\left|V\right|\right)=0$, $V=0$ and
$\left(\mu+\left|V\right|\right)\left|V\right|^{2}=-\frac{27}{2}m\alpha^{2}\Delta_{0}^{2}$.
Considering $\alpha,m,\Delta_{0}\in\mathbb{R}^{+}$ and $\mu,V\in\mathbb{R}$;
we have $D>0$ when $\left(\mu+\left|V\right|\right)\left|V\right|^{2}<-\frac{27}{2}m\alpha^{2}\Delta_{0}^{2}$,
and negative $D$ otherwise. If $D\geq0$ we have three real roots,
otherwise we have one real and two complex roots. A general solution
for differential equation (\ref{eq: qubic Heff diff eq 1}) is $\psi_{1}=C_{1}e^{k_{1}x}+C_{2}e^{k_{2}x}+C_{3}e^{k_{3}x}$,
where we need to enforce the boundary conditions $\psi\left(0\right)=0$
and $\psi\left(\infty\right)=0$. We are working under the constraint $\left|\mu\right|<\left|V\right|$, thus we have that our solutions satisfy $D<0$. If $D<0$ there is one real root, $k_{1}$, and two complex, $k_{2}$
and $k_{3}$. Since $s_{1}$ and $s_{3}$ are real, one finds that $k_{2},k_{3}=-u\pm iv$,
thus $s_{1}=k_{1}-2u$, $s_{2}=-2k_{1}u+\left(u^{2}+v^{2}\right)=0$
and $s_{3}=k_{1}\left(u^{2}+v^{2}\right)$. Therefore:
\begin{itemize}
\item $k_{1},u>0$ or $k_{1},u<0$ are not possible since $s_{2}=0$.
\item $k_{1}<0$ and $u<0$: we have $C_{2}=C_{3}=0$ to satisfy the boundary
conditions at infinity and $C_{1}=0$ to satisfy them at $z=0$. No
solution.
\item $k_{1}>0$ and $u>0$: $C_{1}=0$ to satisfy boundary conditions at
infinity and $C_{2}=-C_{3}$ to satisfy them at $x=0$. Therefore
$\psi_{1}=C_{2}\left(e^{k_{2}x}-e^{k_{3}x}\right)=C_{2}e^{-ux}\sin vx$.
\end{itemize}
Summing up, if there are any Majoranas for $k_{y}=0$, equation \eqref{eq:cubic polynomial}
needs to have a positive real root and two complex roots with negative real part.

For the cubic  polynomial \eqref{eq:cubic polynomial} there is a hyperbolic solution for the real root, $k_1$, given by
\begin{equation}
k_{1}=-2\frac{\left|q\right|}{q}\sqrt{\frac{p}{3}}\cosh\left(\frac{1}{3}\mathrm{arccosh}\left(\frac{3\left|q\right|}{2p}\sqrt{\frac{3}{p}}\right)\right)+\frac{s_{1}}{3},
\end{equation}
where $p$ and $q$ are defined in the main text, in Eqs. \eqref{eq:p} and \eqref{eq:q}. From this equation we can immediately find equations for $u=\frac{-s_{1}+k_{1}}{2}$ and for $v^{2}=\frac{2k_{1}}{u}$. These variables are called $u_d$ and $v_d$ in the main text.

\setcounter{equation}{0}
\renewcommand*{\theequation}{B\arabic{equation}}
%

\section{ Exponential decay of MZMs with static disorder}\label{app:decay}
\begin{figure*}
\subfloat[\label{fig:decay_00}]{%
  \includegraphics[width=0.49\linewidth]{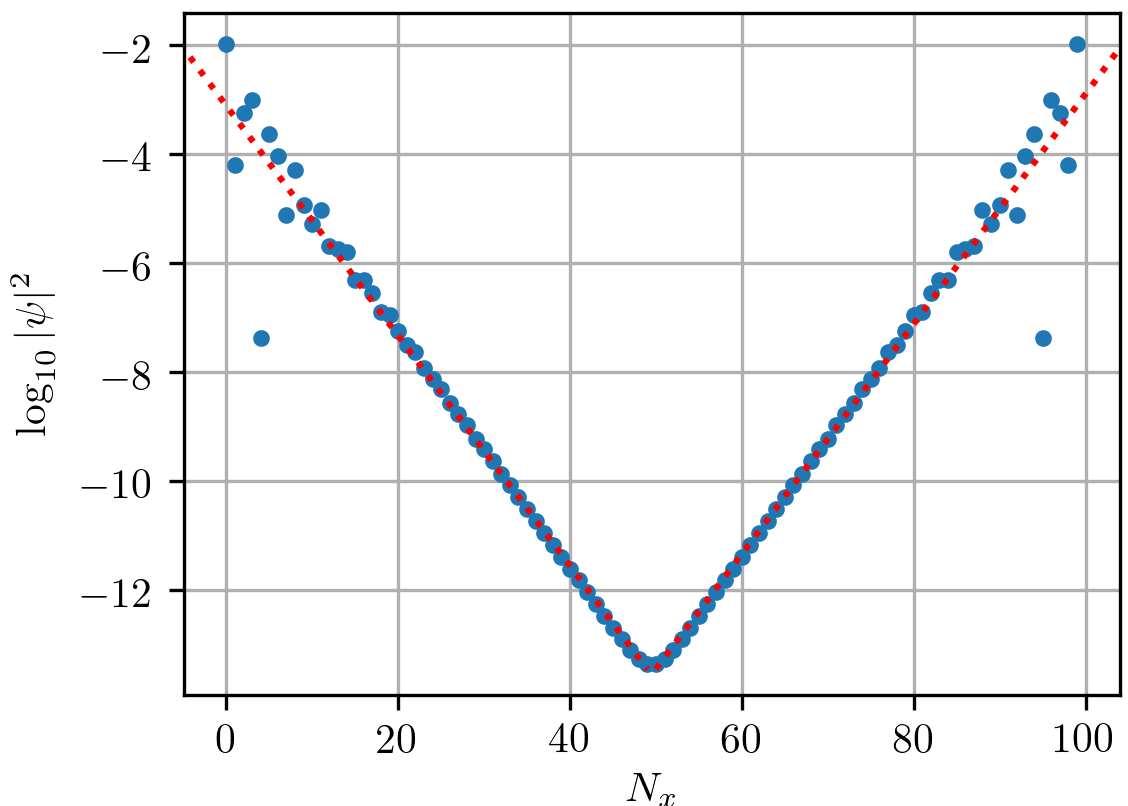}%
}\hfill
\subfloat[\label{fig:decay_01}]{%
 \includegraphics[width=0.49\linewidth]{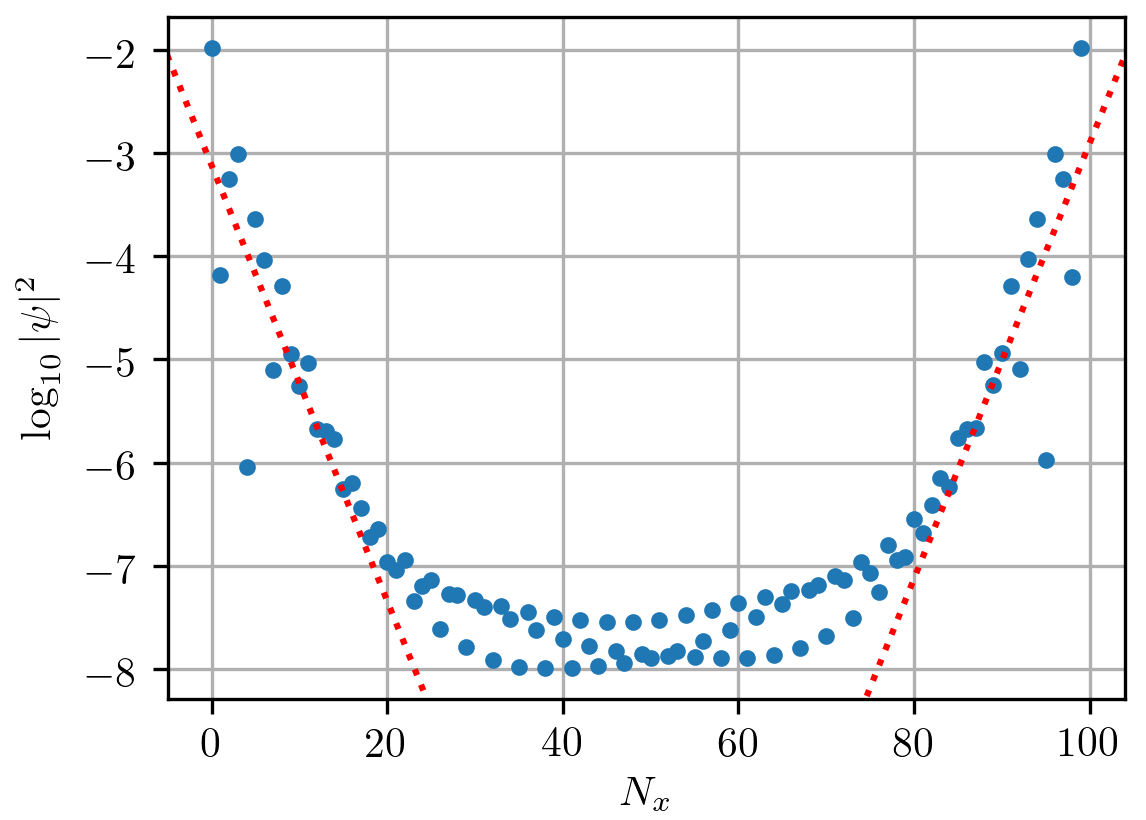}%
}
\caption{Wave function probabilities of the zero energy edge modes on a plane in a logarithmic scale. (a) Shows the wave function of the zero-energy state with no disorder in the system. (b) Depicts the same state with $\sigma_\mu=0.1$.  Parameters: $\tilde{\mu}=2t$, $\tilde{\Delta}=t$. Lattice size is $N_x\times N_y=100\times40$. An average of 15 possible realizations for every 40 possible sections in the $y$-direction was performed.}
\label{fig:Disorder_edge}
\end{figure*}

This appendix is devoted to provide a detailed description of the exponential decay of the MZMs. To this end, we plot the wave function of the zero energy modes in logarithmic scale. If the decay were purely exponential, the wave function would be a straight line. However, we know that there are natural oscillations due to the ansatz of the wave function, Eq. \eqref{eq:majoranas} in the paper.  Fig. \ref{fig:decay_00} shows the decay of a Majorana state without disorder. The red dashed line represents a linear fitting of the results obtained from the lattice model. As it can be concluded from the figure, it is a clearly exponential decay.  The same linear fitting is plotted in Fig. \ref{fig:decay_01}, i.e. the gradient of the red dashed line is the same in both graphics. For weak disorder, the exponential decay remains unaltered.  The fluctuations around the linear fitting shown in Fig. \ref{fig:decay_01}, come not only from the disorder introduced in the system but also from the oscillations of the wave function itself (Eq. \eqref{eq:majoranas}).


We can conclude that the decay of the Majorana modes coming from $d$-wave superconductors remains roughly exponential even when weak static disorder is introduced in the system.
\bibliography{citations}

\end{document}